\documentclass[12pt]{iopart}

\usepackage{iopams}
\usepackage{graphicx}
\usepackage[breaklinks=true, colorlinks=true, linkcolor=blue, urlcolor=blue, citecolor=blue]{hyperref}
\usepackage{dsfont} 
\usepackage{verbatim}
\usepackage{mcite}
\newcommand{\xDownarrow}[1]{%
  \ensuremath{%
    \left\Downarrow\vbox to #1{}\right.\kern-\nulldelimiterspace
  }%
}

\begin{document}
%%%%%%%%
%%%%%%%%
%%%%%%%%
%%%%%%%%
%%%%%%%%
%%%%%%%%

\title[On the Activation of Quantum Nonlocality]{On the Activation of Quantum Nonlocality}

\author{Andr\'es F Ducuara, Javier Madro\~nero and John H Reina} 
\address{Departamento de F\'isica, Universidad del Valle, A.A.
25360, Cali, Colombia}
\address{Centre for Bioinformatics and Photonics---CIBioFi, Calle 13 No. 100-00, Edificio 320-1069, Cali, Colombia}
\ead{andres.ducuara@correounivalle.edu.co
\\ javier.madronero@correounivalle.edu.co
\\ john.reina@correounivalle.edu.co }
\vspace{10pt}
% \begin{indented} \item[]August 2016 \end{indented}

\begin{abstract}
We report on some quantum properties of physical systems, namely, entanglement, nonlocality, $k$-copy nonlocality (superactivation of nonlocality), hidden nonlocality (activation of nonlocality through local filtering) and the activation of nonlocality through tensoring and local filtering. The aim of this work is two-fold. First, we provide a review of the numerical procedures that must be followed in order to calculate the aforementioned properties, in particular, for any two-qubit system, and reproduce the bounds for two-qudit Werner states. Second, we use such numerical tools to calculate new bounds of these properties for two-qudit Isotropic states and two-qubit Hirsch states.
\end{abstract}

% \pacs{ 03.67.-a, 03.65.Ta, 42.50.Lc }

%\noindent{\it Keywords}: Qubits; quantum information; quantum nonlocality; entanglement. \\
% \submitto{\JPA}
%\maketitle

%%%%%%%%%%
%%%%%%%%%%
\section{Introduction}
%%%%%%%%%%
%%%%%%%%%%  

\label{intro}
The understanding and classification of the properties of quantum states are important subjects from both fundamental and practical points of view \cite{E2009, NL2014}. {\emph{Entanglement}} \cite{E2009} and {\emph{nonlocality}} \cite{NL2014} are useful resources for quantum protocols, namely: quantum teleportation \cite{teleportation1993} and cryptography \cite{Ekert1991}. However, a deeper understanding of the relationship between them is still required \cite{NL2014}. Even though entanglement is necessary in order to achieve nonlocality, these two properties are not equivalent, i. e., there exist entangled {\emph{local}} states \cite{Werner1989}. Since several quantum protocols exclusively make use of nonlocality as a resource \cite{NL1}, it is natural to ask whether it is possible to use these (apparently useless) entangled local states in order to achieve (activate) nonlocality by means of processes termed {\emph{activation scenarios}}. There are three of these mechanisms: {\emph{local filtering}} \cite{Popescu1995}, {\emph{tensoring}} \cite{originalactivation2011}, and {\emph{quantum networks}} \cite{networks2011}. Additionally, it is possible to consider combinations of them \cite{NL2014}. 

In this work, we begin by briefly reviewing the aforementioned activation scenarios and pay particular attention to the following: (1) the complete characterisation of {\emph{hidden nonlocality}} (or activation through local filtering) for two-qubit systems recently derived in \cite{PG2015}, (2) a case of the activation through tensoring called {\emph{$k$-copy nonlocality (or superactivation of nonlocality)}}  \cite{superactivation12012}, and (3) activation through the combination of {\emph{tensoring}} and {\emph{local filtering}} \cite{activationbipartite2008, activation2012}. Moreover, we have focused (though not restricted) on the study of these three scenarios over two-qubit systems. The latter could be of interest for future studies on states coming from the dynamics of open quantum systems where entanglement and nonlocality (restricted to the standard definition) are usually investigated \cite{BC1, BC2}. We then used the already mentioned tools to perform numerical simulations in order to investigate these quantum properties for some states of interest. We first reproduced the bounds for the so-called two-qudit Werner states. We then reported new bounds of these properties for the so-called two-qudit Isotropic states and two-qubit Hirsch states.

This work is organised as follows. The first two sections deal with entanglement and nonlocality. The third section establishes the motivation underneath the activation of nonlocality and consequently, we present an overview of the currently known activation scenarios. Next, we address hidden nonlocality, followed by $k$-copy nonlocality, and the activation of nonlocality through tensoring and local filtering.  In the fourth section, we report results regarding the aforementioned quantum properties for some states of interest. First, we reproduced the bounds of these properties for two-qudit Werner states. Second, we report new bounds on the activation through tensoring and local filtering for two-qudit Isotropic states. Third, we report new bounds regarding hidden nonlocality and activation through tensoring and local filtering for two-qubit Hirsch states. Finally, we discuss the obtained results. 

%%%%%%%%%%
%%%%%%%%%%
\section{Entanglement}
%%%%%%%%%%
%%%%%%%%%%

Quantum entanglement was first implicitly introduced in the seminal  1935 EPR article \cite{EPR1935} and subsequent discussions, however, it had to wait until 1989 to be formally defined \cite{Werner1989}. A general finite-dimensional bipartite $AB$ system is represented by a density matrix or quantum state $\rho \in D(\mathds{C}^{d_A}\otimes \mathds{C}^{d_B})$, with $d_A, d_B \geq 2$, where $D(\mathds{H})$ stands for the set of density matrices of the complex Hilbert space $\mathds{H}$ or $D(\mathds{H}):=\{ \rho \in PSD(\mathds{H})|{\rm{Tr}} (\rho)=1\}$, with $PSD$ the set of {\emph{positive semidefinite}} complex matrices, that is the matrices $\rho$ such that $\forall \left| \phi \right> \in \mathds{H}:\left< \phi\right|\rho \left|\phi \right>\geq 0$. A general state there can be written as representing an ensemble of pure quantum states $\{ \left|\psi_i \right>, p_i \}$, with $p_i>0 \hspace{0.2cm}\forall i$, and $\sum_i p_i=1$ as:
\begin{eqnarray}
 \rho=\sum_i p_i  \psi_i,
\label{mixed} 
\end{eqnarray}
with $\psi_i:=\left|\psi_i \right>\left< \psi_i\right|$. We say $ \rho$ is {\emph{separable}} if:
\begin{eqnarray}
 \rho=\sum_i p_i \rho_A\otimes \rho_B,  
\label{entanglement}
\end{eqnarray}
otherwise it is {\emph{entangled}}. Given a quantum state, it is not a trivial task to know whether it is possible to decompose it as in \autoref{entanglement}: there are criteria for entanglement quantification that work quite well for two-qubit systems, and good measures to quantify this are already in place. This said, we lack a unified generalisation to arbitrary high-dimensional and multipartite systems. We next address the main quantifier for two-qubit systems, namely, the {\emph{entanglement of formation}} (EoF) \cite{EoF1996}. The EoF has a compact analytical expression for two qubits \cite{EoF1998} and for a couple of bipartite high-dimensional states we are interested in, namely, the Werner and the Isotropic states \cite{EoFwerner2001, EoFisotropic2000, EoFreview2001}.

%%%%%%%%%%%%%%%%%%%%%
\subsection{Entanglement of Formation (EoF):}
%%%%%%%%%%%%%%%%%%%%%

For pure bipartite states $\psi =  \left|\psi \right>  \left<\psi \right|$, there is an entanglement measure free of ambiguity (in the sense that it is an if and only if criterion) in terms of the well known {\emph{von Neumann entropy}}, $E(\psi):=S(\psi_A)=-{\rm{Tr}}(\psi_A {\rm{log_{2}}}\psi_A)$ with $\psi_A:={\rm{Tr}}_B(\psi)$ the partial trace \cite{EoF1996}. For general mixed states (\autoref{mixed}), it is natural to ask about the possible generalisation $\sum_i p_i E\left(\psi_i \right)$. However, since we have infinite possible ensemble decompositions (\autoref{mixed}), this definition will depend on the chosen ensemble. Consequently, the following measure, known as the \emph{Entanglement of Formation} (EoF), has been introduced \cite{EoF1996}:
\begin{eqnarray}
 E( \rho):= \underset{ \{ \psi_i, p_i \} }{\text{min}}  \left [\sum_i p_i E(\psi_i) \right ].
\label{EoF}
\end{eqnarray}
In 1998, an analytical expression of \autoref{EoF} for the particular case of two-qubit systems was derived, \cite{EoF1998}. Given $\rho \in D(\mathds{C}^2 \otimes \mathds{C}^2) $, its EoF is given by:
\begin{eqnarray}
 EoF( \rho)=h\left( \frac{1+\sqrt{1-C(\rho)^2}}{2}\right),
 \label{EoF22}
\end{eqnarray}
where $h(x):=-x{\rm{log_2}}x-(1-x){\rm{log_2}}(1-x)$ is the so-called binary entropy, $C$ is the so-called concurrence, $C(\rho)=\rm{max}\{0,\lambda_1-\lambda_2-\lambda_3-\lambda_4 \}$, where the $\lambda_i$'s are the square roots of the eigenvalues of the product matrix $\rho \hat \rho$, in decreasing order, with: $\hat \rho= \left(\sigma_y \otimes \sigma_y \right) \rho^* \left( \sigma_y \otimes \sigma_y \right)$, $\rho^*$ the complex conjugate of $\rho$ and, $\sigma_y$ the Pauli matrix. Both, $C(\rho)$ and $EoF(\rho)$ go from $0$ to $1$. There is no known compact analytical expression for general high-dimensional systems, except for a couple of classes of two-qudit states; the so-called Werner and Isotropic states \cite{EoFwerner2001, EoFisotropic2000, EoFreview2001}. Next, we focus on the concept of nonlocality.

%%%%%%%%%
%%%%%%%%%
\section{Nonlocality}
%%%%%%%%%
%%%%%%%%%

\begin{figure}[h!]
 \centerline{\includegraphics[scale=0.3]{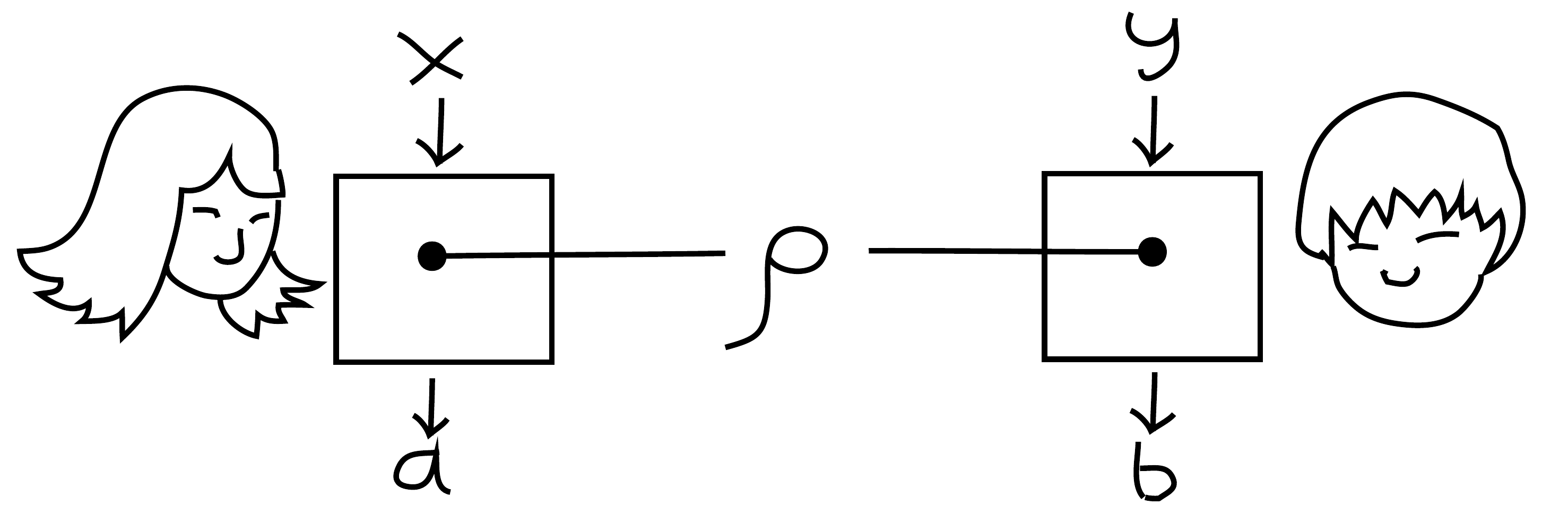}}
\caption{Schematics of a standard Bell test \protect\cite{NL2014}. Experimentalists Alice and Bob, share a bipartite quantum state $\rho$. They make measurements $x$ and $y$ and obtain outcomes $a$ and $b$ respectively.}
\label{fig:fig1}
\end{figure}

Quantum nonlocality was first implicitly introduced in the seminal EPR article \cite{EPR1935} and corresponding subsequent discussions. However, it had to wait until 1964 in order to be formally defined by J. S. Bell \cite{bell1964}. We proceed as follows: Given a bipartite system $\rho \in D(\mathds{C}^{d_A} \otimes \mathds{C}^{d_B})$, the first party of the whole system is sent to experimentalist A (Alice), and the second to experimentalist B (Bob). They want to  study the two observables $A$ and $B$ respectively. These can be  written in their spectral decomposition as $\nonumber A= \sum_a o_aP_a^x, \hspace{0.2cm} B=\sum_b o_bP_b^y$, where: $M^x:= \{ P_a^x \}$ and $M^y:= \{ P_b^y \}$ are sets of projections, and $o_a, o_b$ their respective eigenvalue spectra. However, they can also be written more generally as $\nonumber A= \sum_a o'_aE_a^x, \hspace{0.2cm} B=\sum_b o'_bE_b^y$, where: $M^x:= \{ E_a^x \}$ and $M^y:= \{ E_b^y \}$ are now Positive Operator Valued Measurements (POVM's) with $o'_a, o'_b$ real numbers though not necessarily eigenvalues of $A$ and $B$. Using this quantum state $\rho$, Alice and Bob make measurements $x$ and $y$, obtaining outcomes $a$ and $b$. After repeating the experiment enough times, eventually, they are able to establish the probability of obtaining outcomes $a,b$ after measuring $x$ and $y$. This can be seen as the joint conditional probability function $p(a,b|x,y,\rho)$. According to quantum mechanics, we could reproduce this statistics by means of the Born Rule \cite{NC2010}:
\begin{eqnarray}
 p_{Q}(a,b|x,y,\rho):={\rm{Tr}}\left [ (E_a^x \otimes E_b^y )\rho \right ].
\label{quantum}
\end{eqnarray}
Since quantum mechanics has been proven correct so far, we have: $p(a,b|x,y,\rho)=p_{Q}(a,b|x,y,\rho)$. We can see a sketch of this process in \autoref{fig:fig1}.
Bell introduced the notion of locality \cite{bell1964} within a formalism covering theories that could reproduce this very same statistics (\autoref{quantum}). In this formalism there are probability functions $\mu$ and $\xi$, such that
\begin{eqnarray}
p_{L}(a,b|x,y,\rho)=\int_{\Lambda} d \lambda \mu(\lambda|\rho)\xi(a,b|x, y, \lambda),
\label{local}
\end{eqnarray}
where $\Lambda$ is the often-called {\emph{set of hidden variables}} and the triplet ($\Lambda, \mu, \xi$) is the {\emph{ontological model}}. The {\emph{locality}} condition reads $\xi(a,b|x, y, \lambda)=\xi_A(a|x, \lambda)\xi_B(b|y,\lambda)$. Mathematically speaking, it means that the function $\xi$ is probabilistically independent from sides $A$ and $B$. Physically speaking, it means that Alice and Bob can reproduce the statistics of \autoref{quantum} with the local hidden variable model ($\Lambda, \mu, \xi_A, \xi_B$). Then, the question that arises is the following: Is it possible to reproduce the statistics given by \autoref{quantum} by means of \autoref{local}? i. e.,
\begin{eqnarray}
\nonumber p_{Q}(a,b|x,y,\rho)\stackrel{?}{=}p_{L}(a,b|x,y,\rho).
\end{eqnarray} % that after receiving inputs $x$ and $y$, Alice cannot influence Bob's output and vice versa.
If this is indeed possible for any POVM (we could relax the condition to just projections), the quantum state $\rho$ will be called {\emph{local}}, otherwise $\rho$ will be {\emph{nonlocal}}. For instance, it is not hard to prove that separable states (\autoref{entanglement}) are local states. From the locality condition (\autoref{local}), it is possible to derive the so-called Bell inequalities \cite{bell1964, NL2014} such that, if a quantum state violates one of those inequalities, then that state is nonlocal. In addition to having introduced this concept, Bell also derived the very first Bell inequality with which he was capable of proving the nonlocality of the two-qubit state $\left| \psi \right >:=\frac{1}{\sqrt{2}}\left(\left| 01\right>- \left| 10\right>\right)$ \cite{bell1964}. It is well known that nonlocality and entanglement are \emph{equivalent} for multipartite pure states (in the sense that every entangled state is nonlocal and viceversa) \cite{Gisin1991, GP1992, PR1992}. However, this is not the case for general mixed states, i. e., there exist mixed entangled local states \cite{Werner1989, NL2014}. For a bipartite system, the aforementioned Bell inequalities can be classified through the parameters $(m_A, m_B, n_A, n_B)$ where $x=1,...,m_A$, $y=1,...,m_B$, $a=1,...,n_A$, and $b=1,...,n_B$ \cite{NL2014}. Here, we shall discuss the simplest non trivial case, $(2,2,2,2)$. This inequality has a very compact form, at least for two-qubit systems, which we detail in what follows.

%%%%%%%%%%%%%%
\subsection{CHSH Inequality:}
%%%%%%%%%%%%%%

For a bipartite system, the Clauser-Horne-Shimony-Holt (CHSH) inequality \cite{CHSH1969} considers two dichotomic (eigenvalues $\pm 1$) observables per party, namely, ($A_1, A_2, B_1, B_2$). With the notation already introduced, this would be the inequality (2,2,2,2) and it takes the form:
\begin{eqnarray}
\left|B_\rho(A_1, A_2, B_1, B_2)\right|:=\left |E_{11}+E_{12}+E_{21}-E_{22}\right|\leq 2,
\label{CHSH}
\end{eqnarray}
where $E_{ij}:= {\rm{Tr}}\left [\left(A_i\otimes B_j\right)\rho \right]$, $i,j=1,2$. Both, this and more general Bell inequalities can be derived in a systematic way by means of a geometric approach \cite{NL2014}. The idea is to maximise the $B_\rho$ function over those four observables. There is no compact solution for general arbitrary dimensions, except for two qubits \cite{HHH1995}. In this case  any state $\rho \in D(\mathds{C}^2 \otimes \mathds{C}^2)$, can be written as:
\begin{eqnarray}
\nonumber \frac{1}{4}\left (\mathds{1}_{4\times4} +   \vec \sigma \cdot \vec a  \otimes \mathds{1} +\mathds{1} \otimes \vec \sigma \cdot  \vec b+\sum_{n,m=1}^3 t_{n,m}\sigma_n \otimes \sigma_m \right ),
\end{eqnarray}
with $\vec \sigma=[\sigma_i]$, $\sigma_i$, $i=1,2,3$, the Pauli matrices, $\vec a=[a_i]$ with $ a_i:= {\rm Tr} [(\sigma_i \otimes  \mathds{1})\rho]$, $\vec b=[b_i]$ with $b_i:= {\rm Tr} [(\mathds{1}\otimes \sigma_i)\rho]$, and $t_{nm}:={\rm{Tr}}[\rho (\sigma_n  \otimes \sigma_m)]$ making the matrix $T_{\rho}:=[t_{nm}]\in M_{3\times3}(\mathds{R}) $. We have the following characterisation in terms of the sum $M(\rho):=\mu+\widetilde \mu$ of the two biggest eigenvalues $\mu, \widetilde \mu$ of the matrix $U_\rho:=T_{\rho}^TT_{\rho}\in M_{3\times3}(\mathds{R}) $:
\begin{eqnarray}
\nonumber \rho \hspace{0.3cm} \text{violates} \hspace{5.5cm} \\
\text{the CHSH } \hspace{1.3cm} \xLeftrightarrow{\hspace{1cm}} \hspace{0.3cm} M(\rho):=\mu+\widetilde \mu>1\,.\label{criterion}\\
\nonumber \text{Inequality (\autoref{CHSH}) \hspace{4.8cm}}
\end{eqnarray}
This is because, it is possible to show that $\rm{max} B_\rho:=| \rm{max}_{A_1, A_2, B_1, B_2}B_\rho|=2\sqrt{M(\rho)}$. Then, using the Tsirelson's bound $\rm{max } B_\rho\leq 2\sqrt{2}$ \cite{Tsirelson1980}, it follows $0\leq M(\rho)\leq2$, showing nonlocality in the interval $1<M(\rho)\leq 2$. Instead of $M(\rho)$, we could work with $B(\rho):=\sqrt{ \rm{max}\left \{0, M(\rho)-1\right \}}$ because for pure states we have that the former turns out to be equal to the concurrence already discussed, i. e., $C(\left| \psi \right>)=B(\left| \psi \right>)$ \cite{equivalence2004}. However, in order to analyse nonlocality through the CHSH inequality and make a direct comparison with EoF, we will plot:
\begin{eqnarray}
 CHSH(\rho)=h\left( \frac{1+\sqrt{1-B(\rho)^2}}{2}\right),
 \label{CHSH22}
\end{eqnarray}  
being $h$ the binary entropy defined in the previous section. It should be pointed out that, if $\rho$ does not violate the CHSH inequality, it does not necessarily imply that $\rho$ is a local state; actually, it is possible that the state violates another inequality, for instance the state considered in \cite{CG2004}. On the other hand, proving locality is perhaps trickier than doing so for nonlocality \cite{NL2014}. To wrap up, we search for nonlocality by means of the CHSH inequality, however, there are entangled states that do not violate it and even more, there are entangled local states (states that do not violate any Bell inequality). Let us address the process of how to activate nonlocality on these last classes of states.

%%%%%%%%%%%%%%%%%%%%
%%%%%%%%%%%%%%%%%%%%
\section{Activation of Nonlocality: Scenarios}
%%%%%%%%%%%%%%%%%%%%
%%%%%%%%%%%%%%%%%%%% 

On the one hand, we have already pointed out that, although for pure states entanglement and nonlocality are equivalent, (in the sense that every entangled state is nonlocal and vice versa) this is not the case for general mixed states. This, in principle, would settle the question about the relationship between entanglement and nonlocality. However, in the mid nineties, Popescu was able to activate the {\emph{hidden nonlocality}} of a class of entangled local states after a proper manipulation of them \cite{Popescu1995}.  This raised again the question about the relationship between entanglement and what we could call, new definitions (generalisations) of nonlocality. Therefore, from a fundamental point of view, it is interesting to explore the aforementioned new relationship.

On the other hand, and from a practical point of view, even though entanglement has been regarded as resource for many tasks \cite{E2009}, it is also well known that there are protocols that explicitly require either nonlocality or the violation of certain Bell inequalities \cite{NL2014, NL1}. Let us consider the following situation: suppose that an experimentalist is able to prepare entangled local states only, but she needs to implement protocols that require nonlocality. What could he do about it? These procedures are called {\emph{activation scenarios}} and we address them in what follows \cite{NL2014}:

\begin{figure}[h!]
 \centerline{\includegraphics[scale=0.27]{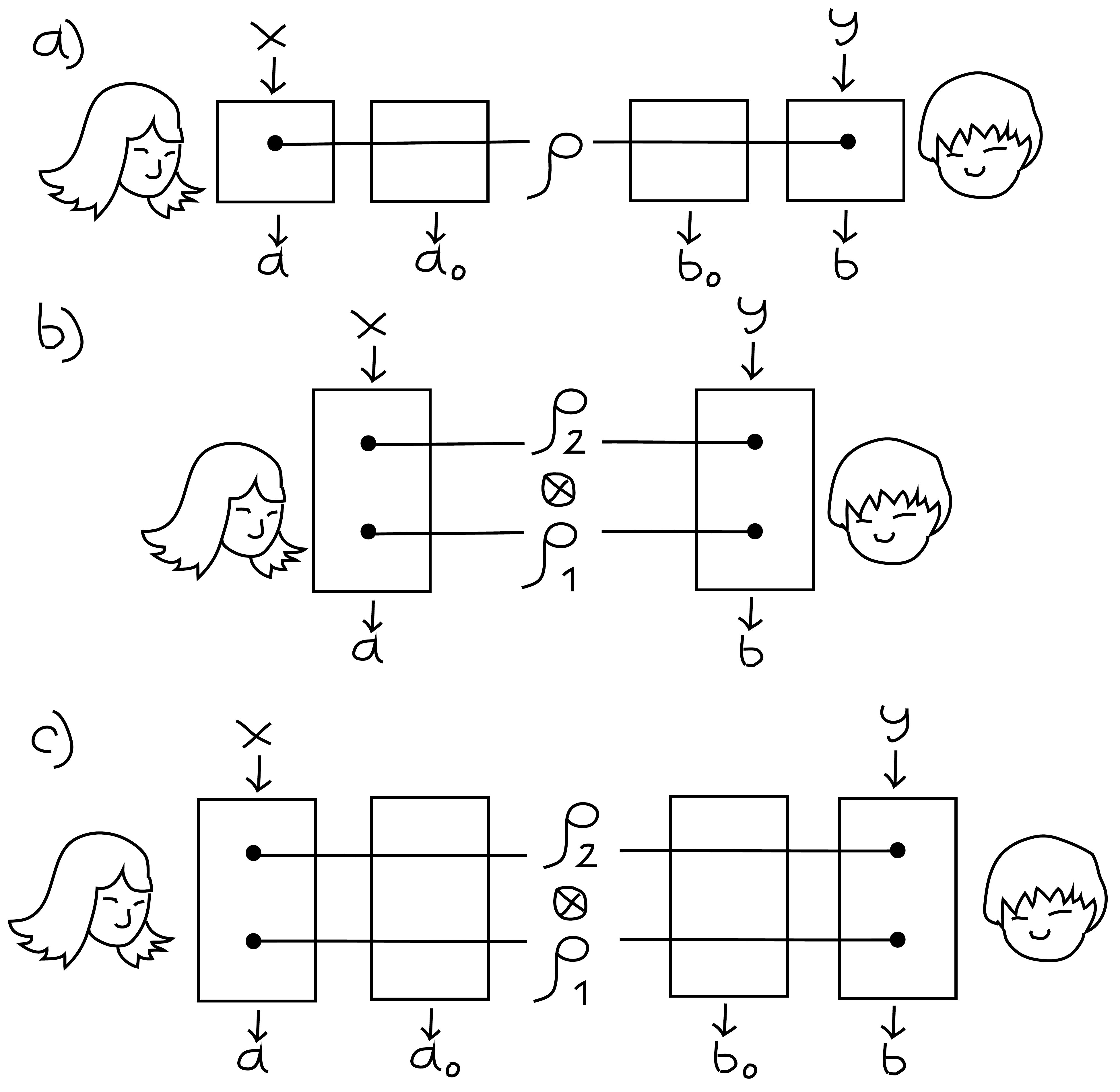}}
\caption{Activation of nonlocality through Local Filtering \protect\cite{Popescu1995}. Experimentalists Alice and Bob receive a part of a bipartite quantum state $\rho$. They make a local filtering (map onto a two-qubit system) before a standard Bell test. {\bf{(b)}} Activation of nonlocality through Tensoring \protect\cite{originalactivation2011}.  Experimentalists Alice and Bob receive a part of a quantum state $\rho_1$ and part of another quantum state $\rho_2$, then, they make a standard Bell test. {\bf{(c)}} Activation of nonlocality through the combination of Tensoring and Local Filtering.}
\label{fig:fig2}
\end{figure}

%%%%%%%%%%%%%%%%%%%%%%%
\subsection{Activation through Local Filtering (LF):} 

Popescu in Ref. \cite{Popescu1995} took Werner entangled local states $\rho_W \in D(\mathds{C}^d \otimes \mathds{C}^d)$ and after applying a {\emph{local filter}}, an operation that could be thought as a projection onto a two-qubit system, namely $P \otimes Q$ with $P:=|1\left>\right<1|+|2\left>\right<2|$, and $Q:=|1\left>\right<1|+|2\left>\right<2|$, he found that the final state violated the CHSH inequality (\autoref{CHSH}), with $d>5$ (See Appendix A for a detailed explanation). We can see a sketch of the procedure in \autoref{fig:fig2} (a). Soon after this result, Gisin \cite{gisin1996} proposed another example, but this time with states that, even though it was not known whether they were local, at least, they did not violate the CHSH inequality. Recently, Pal \& Ghosh \cite{PG2015} derived a complete characterisation for two-qubit systems.
 
%%%%%%%%%%%%%%%%%%%%%
\subsection{Activation through Tensoring (T):} 
 
 Here, the question is: Is it possible to find $\rho_1$ and $\rho_2$ entangled local states such that $\rho_1 \otimes \rho_2$ is an entangled nonlocal state? This procedure is illustrated in \autoref{fig:fig2} (b). If $\rho_2=\rho_1$, or in general if $\rho \otimes ^k$ is nonlocal, the phenomenon is called {\emph{superactivation}}. Similarly, we could also say that the state $\rho$ is {\emph{$k$-copy nonlocal}} or that we have activation through {\emph{$k$-tensoring}}. It is also natural to think about the combination of this procedure with the previous one as we can see in \autoref{fig:fig2} (c). Historically, the first tensoring procedure appeared as a combination of $k$-tensoring with LF \cite{Peres1996, Masanes2006}. Then, general tensoring and LF would arise \cite{activationbipartite2008, activation2012}. The protocol of activation through tensoring alone (without LF) was reported in \cite{originalactivation2011}, later, superactivation was put forward \cite{superactivation12012}. 
 
We next address the following three scenarios. First, hidden nonlocality \cite{PG2015}. Second, $k$-copy nonlocality \cite{superactivation12012}. Third, tensoring with local filtering (first tensoring, then local filtering) \cite{activationbipartite2008, activation2012}. We shall focus on numerical approaches for two-qubit systems.

%%%%%%%%%%%%%%%%%%%%%%
%%%%%%%%%%%%%%%%%%%%%%
\section{Local Filtering: Hidden Nonlocality (HN)} 
%%%%%%%%%%%%%%%%%%%%%%
%%%%%%%%%%%%%%%%%%%%%%

Recently, Pal \& Ghosh \cite{PG2015} derived a complete characterisation of hidden nonlocality for two-qubit systems. Given $\rho \in D(\mathds{C}^2 \otimes \mathds{C}^2)$, $\rho$ possesses hidden nonlocality (or $\rho$ can be transformed into a CHSH-violating state $\rho'$ by means of local filters) if and only if $\lambda_\rho^1+\lambda_\rho^2>\lambda_\rho^0$, where the $\lambda_\rho^i$'s are the eigenvalues of the matrix $C_{\rho}:=\eta T_{\rho}\eta T_{\rho}^T$ organized in decreasing order, with $\eta:={\rm diag}(1,-1,-1,-1)$ and $T_{\rho}:=[t_{nm}]$ a matrix with elements $t_{nm}:={\rm Tr}[(\sigma_m \otimes \sigma_n)\rho]$ as defined in the CHSH nonlocality section. The CHSH-violation of that new state $\rho'$ is given by $M'(\rho):=M(\rho')=\frac{\lambda_\rho^1+\lambda_\rho^2 }{\lambda_\rho^0}$. Instead of $M'(\rho)$, we could work with $B'(\rho):=\sqrt{ \rm{max}\left \{0, M'(\rho)-1\right \}}$ because for pure states we have that the former turns out to be equal to the concurrence already discussed, i. e., $C(\left| \psi \right>)=B(\left| \psi \right>)$ \cite{equivalence2004}. Then, in order to analyse hidden nonlocality, we will plot:
\begin{eqnarray}
HN(\rho)=h\left( \frac{1+\sqrt{1-B'(\rho)^2}}{2}\right),
\label{HN22}
\end{eqnarray}     
being $h$ the binary entropy defined in the EoF and CHSH sections.

%%%%%%%%%%%%%%%%%%%%%%%%%%%
%%%%%%%%%%%%%%%%%%%%%%%%%%%
\section{ Tensoring: $k$-copy Nonlocality \\ (Superactivation of Nonlocality SA)}
%%%%%%%%%%%%%%%%%%%%%%%%%%%
%%%%%%%%%%%%%%%%%%%%%%%%%%%

In this section we focus on the case of the activation of nonlocality only through tensor product; in particular, when the state is capable of activating nonlocality by itself, i. e., superactivating \cite{superactivation12012}. We shall specialise our results on two-qubit systems.

%%%%%%%%%%%%%
\subsection{Main Theorems:}

In order to talk about $k$-copy nonlocality, the teleportation protocol \cite{teleportation1993} will come in handy. Given a two-qubit state $\rho \in D(\mathds{C}^2\otimes \mathds{C}^2)$,  we have the next hierarchy of properties depicted by the following chain of implications:

\vspace{0.5cm}

\hspace{-0.4cm} $\rho$ violates the CHSH inequality (\autoref{CHSH}).

\hspace{1cm}\xDownarrow{0.5cm} \cite{HHH1996}

\hspace{-0.4cm}$\rho$ is useful for teleportation.

\hspace{1cm}\xDownarrow{0.5cm} \cite{superactivation22013})

\hspace{-0.4cm}$\rho$ is $k$-copy nonlocal.

\vspace{-2.5cm}

\begin{eqnarray}
\phantom{asdfasdfasdf}
\label{implications}
\end{eqnarray}

\vspace{1.5cm}

The proofs of both implications can be found in \cite{HHH1996}, \cite{superactivation22013}, respectively. Here, we comment on a couple of remarks: The first implication is restrictive in the sense that there exist entangled  local states (states that do not violate any Bell inequality, in particular, the CHSH inequality), although useful for teleportation \cite{Popescu1994}. The second implication actually holds true even for general two-qudit systems $\rho \in D(\mathds{C}^d\otimes \mathds{C}^d)$ \cite{superactivation22013}. Third, usefulness for teleportation does not cover all the entangled states set, i. e., there exist entangled states not useful for teleportation \cite{HHH1996}. Fourth, it remains open both, the existence of an entangled never $k$-copy nonlocal state and a $k$-copy nonlocal, not useful for teleportation state. Finally, since usefulness for teleportation is a property which is possible to search for numerically, it consequently allows us to enquire about $k$-copy nonlocality. In what follows, we discuss how to numerically calculate usefulness for teleportation.

The teleportation protocol \cite{teleportation1993}, in a nutshell, works as follows. A quantum pure state $\left| \phi \right> \in \mathds{C}^d$ can be teleported by means of a channel built with another quantum state $\rho \in D(\mathds{C}^d\otimes \mathds{C}^d)$. In order to check how useful $\rho$ is to the protocol, a function called {\emph{Fidelity of Teleportation}} (FoT) was proposed \cite{Popescu1994}. A seminal result regarding this function \cite{HHH1999} establishes that:
\begin{eqnarray}
 F_d(\rho)=\frac{df_d(\rho)+1}{d+1},
\label{FoT}
\end{eqnarray}
where $f_d$ is the so called {\emph{Fully Entangled Fraction}} (FEF) (sometimes also called {\emph{Entanglement Singlet Fraction}}) given by:
\begin{eqnarray}
 f_d(\rho):= \underset{ \psi \in ME  }{\rm{max}} \left<\psi| \rho| \psi \right>,
\label{FEF}
\end{eqnarray}
where $ME$ stands for the set of maximally entangled states, i. e., the states $\psi:=\left| \psi \right>\left<\psi \right|$, defined as: $\left| \psi \right>:=\left( U_1 \otimes U_2 \right) \frac{1}{\sqrt{d}}\sum_{i=0}^{d-1} \left|ii \right>$, with $U_1, U_2$ local unitary operations. Taking a look at \autoref{FEF}, we have $\frac{1}{d+1}<F_d<1$. If we use separable states in the FEF (\autoref{FEF}), we obtain the value  $f_d(\rho_{sep})=\frac{1}{d}$ \cite{HHH1999}. Consequently, we have the characterisation:
\begin{eqnarray}
\nonumber \rho \hspace{0.2cm} \text{is useful for teleportation} \hspace{0.3cm} \xLeftrightarrow{\hspace{0.8cm}} \hspace{0.3cm} f_d(\rho)>\frac{1}{d}.
\end{eqnarray}
In particular, for a two-qubit state $\rho$ ($d=2$), with $\frac{1}{3}<F_d(\rho)<1$, $\rho$ is useful for teleportation if and only if $f_2(\rho)>\frac{1}{2}$ or $F_2(\rho)>\frac{2}{3}$. Therefore, we can search for $k$-copy nonlocality through the usefulness for teleportation by means of the calculation of the FoT (\autoref{FoT}). A natural question that arises is: what is the actual $k$ value necessary for this superactivation? It turns out that, from the second implication in \autoref{implications}, it is possible to extract this value \cite{superactivation12012, superactivation22013}. This $k$ value is a minimum because it is necessary only one nonlocal state in order to activate nonlocality through tensoring \cite{PR1992}. We next show how to calculate this $k$.

%%%%%%%%%%%%%%%%%%%%%%%%%%%
\subsection{The superactivation tensoring factor $k(d,f_d)$:}

\begin{figure}[h!]
 \centerline{\includegraphics[height=3.7in,width=3.1in]{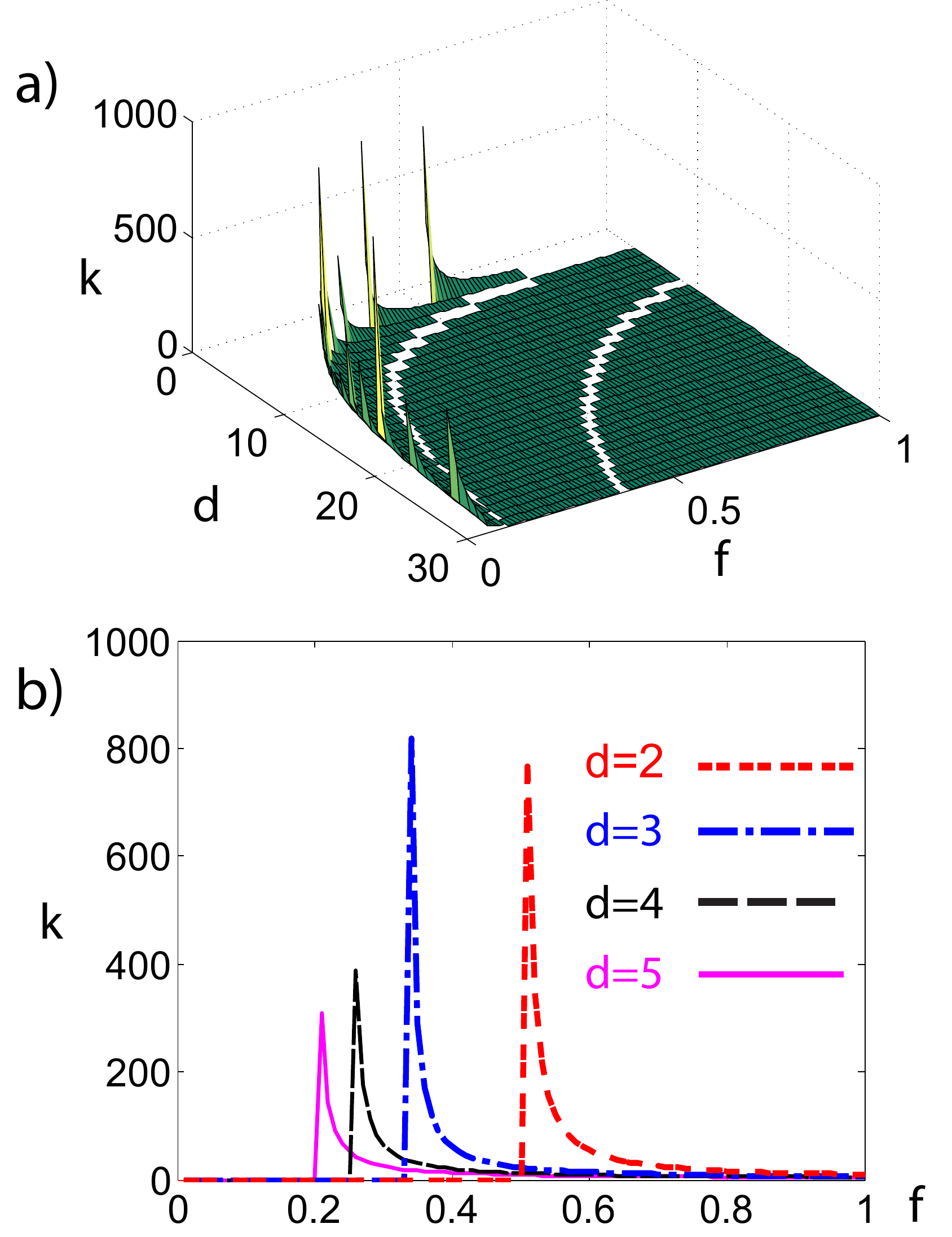}}
\label{fig:fig3}
\caption{{\bf{(a)}} Superactivation tensoring factor $k(d, f_d)$ in terms of the fully entangled fraction (FEF) $0\leq f_d\leq1$ and the dimension $d \geq 2$ of the two-qudit system. The white curves represent the threshold for $k<10$ and $k<3$ from left to right respectively. {\bf{(b)}} Transversal cuts of \autoref{fig:fig3} (a) for $d=2,3,4$ and $5$.}
\end{figure}

By means of the above-mentioned theorem, the second implication in \autoref{implications} it is possible to extract the explicit values of $k$ in terms of the dimension $d$ of the two-qudit system and their fully entangled fraction $f_d$. From the theorem, those $k$'s must satisfy the following relation:
\begin{eqnarray}
  \left [ \frac{C' }{C( \ln d)^2} \right ]\frac{\left (f_d d \right)^k}{k^2}>1,
   \label{anterior}
\end{eqnarray}
with constants $C:=e^4, \hspace{0.2cm}C' :=4$. Since in  \autoref{anterior}, we have a function to the power of $k$ over a polinomial (quadratic) function of $k$, we see that the activation goes from $f>1/d$. Numerically, in \autoref{fig:fig3} (a), we see the behaviour of $k(d,f_d)$. For most of the $d-f$ region, $k<10$ is enough (as it is depicted by the white curve in the left); just when $f_d$ is close to the boundary $\frac{1}{d}$, it becomes asymptotically more difficult to superactivate it. In \autoref{fig:fig3} (b), we see cuts for $d=2,3,4$ and $5$. Next, we will analyse how to calculate $f_d$ for two qubits ($d=2$) \cite{inclusive2002}.

%%%%%%%%%%%%%%%%%%%%
\subsection{Fidelity of Teleportation (FoT):}

Following \cite{inclusive2002}, we summarise the Fidelity of Teleportation (FoT) characterisation for two-qubit systems. Given $\rho \in D(\mathds{C}^2 \otimes \mathds{C}^2)$, we have that $f_2(\rho)={\rm{max}}\{\eta_{i}, 0\}$, where the $\eta_{i}$'s are the eigenvalues of the matrix $M=[M_{mn}]$, with elements $M_{mn}= {\rm{Re}} \left (\left<\psi_m \right| \rho \left|\psi_n \right> \right)$, and $\{\left|\psi_n \right>\}$ the so-called magic basis $\left| \psi_{ab}\right>:=i^{(a+b)}(\left| 0,b\right>+(-1)^a\left|1,1\oplus b \right>)/\sqrt{2}$. The FoT is given by  $F_2(\rho)=\frac{2f_2(\rho)+1}{3}$, and we have already seen that $\rho$ is useful for teleportation if and only if $f_2>\frac{1}{2}$ or $F_2(\rho)>\frac{2}{3}$. It should be pointed out that, the bound $f_2=\frac{1}{2}$ is also a measure of usefulness for other two-qubit protocols \cite{inclusive2002}. In order to enquire about this property, we shall plot
\begin{eqnarray}
SA(\rho)=F_2'^+ ,\hspace{0.3cm} \text{with} \hspace{0.3cm} F_2':=F_2-\frac{2}{3},
\label{SA22}
\end{eqnarray}
and $^+$ the positive part of the function. For multipartite or high-dimensional systems, there is not a general explicit formula for the FEF, except for Werner and Isotropic states \cite{FEFwi2010}.

%%%%%%%%%%%%%%%%%%%%%
%%%%%%%%%%%%%%%%%%%%%
\section{Tensoring and Local Filtering (T\&LF)}
%%%%%%%%%%%%%%%%%%%%%
%%%%%%%%%%%%%%%%%%%%%

In this section, we deal with the activation of nonlocality through tensoring and local filtering \cite{activationbipartite2008, activation2012}. Let us consider the set $P_{{\rm{CHSH}}}$ formed by states satisfying the CHSH inequality even after all possible local filtering (LF) operations \cite{activationbipartite2008}. In other words: $P_{\rm CHSH}\subset D(\mathds{H})$,
\begin{eqnarray}
\nonumber \rho \in P_{{\rm{CHSH}}} \hspace{0.2cm} \xLeftrightarrow{\hspace{0.8cm}} \hspace{0.2cm} \forall \hspace{0.0cm} \Omega: D\left ( \mathds{H} \right) \longrightarrow D\left (\mathds{C}^2\otimes \mathds{C}^2 \right) \hspace{0.1cm} \text{LF}  
\end{eqnarray}
\vspace{-0.5cm}
\begin{eqnarray}
\nonumber \text{ it holds that, $\Omega \left(\rho \right)$ does not violate CHSH}.
\end{eqnarray}
Here, the local filter LF denotes a \emph{separable map} of the form $\Omega (\rho) = \sum_i (A_i \otimes B_i) \rho (A_i \otimes B_i)^{\dagger}$, with $A_i, B_i$ being Kraus operators \cite{activationbipartite2008}. This $P_{\rm CHSH}$ set has been characterised for two-qubit systems \cite{ref30}. If we define $P'_{\rm CHSH}$ as the set of states that do not violate CHSH even after $k$ tensoring themselves or LF, also called {\emph{not asymptotically violation}} \cite{Masanes2006}, we have the relation $P'_{\rm CHSH} \subset P_{{\rm{CHSH}}}$. There is an equivalence between this asymptotically violation of CHSH \cite{Masanes2006} and another property called {\emph{distillability}} \cite{distillability1996}, which will let us search numerically for states in $P_{{\rm{CHSH}}}$. 

%%%%%%%%%%%%%
\subsection{Main Theorem:}

Since we are focused on two-qubit systems, we highlight the result for bipartite systems \cite{activationbipartite2008}. However, it should be said that the theorem also works for general multipartite systems \cite{activation2012}. It establishes the following:
\begin{eqnarray}
\nonumber \tau  \in D\left( \mathds{C}^{d_A}\otimes \mathds{C}^{d_B}\right)  \hspace{0.7cm} \exists  \rho_\tau \in D\left[ \bigotimes_{i=1}^2 \left(\mathds{C}^{d_i}\otimes \mathds{C}^2\right)\right]  \hspace{-0.5cm}\\
\hspace{0.7cm}   \text{is} \hspace{1.4cm}\xLeftrightarrow{\hspace{1cm}} \hspace{0.7cm}  \text{s.t.} \hspace{0.2cm}\rho_\tau \in P_{\rm CHSH},   \label{TandLF}\\
\nonumber  \text{entangled} \hspace{2.5cm}\tau \otimes \rho_\tau \notin P_{\rm CHSH}.
\end{eqnarray}
A couple of remarks: first, since the result holds for any entangled state $\tau$, it also applies for entangled states $\tau \in P_{\rm CHSH}$, which would imply that the CHSH nonlocality of  $\tau$ has been activated (in the sense of $P_{\rm CHSH}$) through tensoring (with $\rho_\tau$) and LF. We could also call it tensorial activation of hidden nonlocality. Second, even though the theorem guarantees the existence of the matrix $\rho_\tau$, it does not tell us explicitly how to calculate it. 

%%%%%%%%%%%%%%%%
\subsection{Numerical Approach:}

Given $\tau \in D(\mathds{C}^{d_A} \otimes \mathds{C}^{d_B})$ an entangled state, we would like to find the aforementioned respective density matrix $\rho_\tau$. In order to do so, we follow the approach reported in \cite{activation2012}. From the proof of the main theorem, which can be found in \cite{activation2012}, it is enough to look for a density matrix  $\rho_\tau \in D\left[(\mathds{C}^{d_A} \otimes \mathds{C}^{2})\otimes (\mathds{C}^{d_B}\otimes \mathds{C}^{2})\right]$ with the following characteristics. First,
\begin{eqnarray}
 \rm{Tr} \left[\rho_{\tau} \left(\tau^{\rm{T}} \otimes H_{\pi/4} \right)\right]<0, 
 \label{numerical}
\end{eqnarray}
\vspace{-0.5cm}
\begin{eqnarray}
\nonumber \text{with:} \hspace{0.5cm} H_{\theta}:=\mathds{1}\otimes\mathds{1}-\cos \theta \sigma_{\rm x} \otimes \sigma_{\rm x}-\sin \theta \sigma_{\rm z} \otimes \sigma_{\rm z} .
\end{eqnarray}
Second, we have to check that $\rho_\tau \in P_{\rm CHSH}$. In principle, it is not numerically possible to check if a state belongs or not to the set $P_{\rm CHSH}$ (with the exception of two-qubit systems \cite{PG2015}, however, our matrix $\rho_\tau$ is not of this sort). Fortunately, we have some results that partially allow us to search for it: i) as we have already pointed out, $P'_{\rm CHSH} \subset P_{\rm CHSH}$; ii) $P'_{\rm CHSH}$ is equivalent to {\emph{Bound Entanglement}} ($BE$) \cite{Masanes2006}; iii) there is a class of these bound entangled states, the so-called $PPT$ (Positive Partial Transpose) states \cite{PPT1996, PPTinBE1996}. The former are states such that, their partial transpose respect to the first subsystem is positive or $\rho^{T_1}\geq 0$. In other words, given $\rho \in D(\mathds{C}^{d_A} \otimes \mathds{C}^{d_B})$, which is always possible to write it as $\rho:=\sum_{ijkl} \rho_{ijkl} |i\left>\right<j |\otimes |k\left>\right<l|$ with $\rho_{ijkl}$ coefficients, the partial transpose respect to the first subsystem is defined as:
\begin{eqnarray}
\nonumber \rho^{T_1}&:= & \sum_{ijkl} \rho_{ijkl} (|i\left>\right<j |)^T\otimes |k\left>\right<l| \\
\nonumber &=&  \sum_{ijkl} \rho_{ijkl} |j\left>\right<i|\otimes |k\left>\right<l|,
\end{eqnarray}
 with $T$ the standard transposition. The state $\rho$ is then a $PPT$ state, if its partial transpose remains positive or $\rho_{\tau} ^{T_1}\geq 0$. Therefore, we have the hierarchy of properties: 
\begin{eqnarray}
\nonumber PPT \longrightarrow BE \longleftrightarrow P'_{\rm CHSH}\longrightarrow P_{\rm CHSH}. 
\end{eqnarray}
Hence, we could look  for a matrix $\rho_\tau$ with a positive partial transposition respect to the first subsystem $\mathds{C}^{d_A} \otimes \mathds{C}^{2}$ in order to guarantee that it belongs to $P_{\rm CHSH}$. Then, we have the following minimisation problem: 
\begin{eqnarray}
 \text{minimise:}\hspace{0.6cm} \sigma(\tau):=\rm{Tr} \left[\rho_{\tau} \left(\tau^{\emph{T}} \otimes H_{\pi/4} \right)\right],   \label{minimisation}
 \end{eqnarray}
 \vspace{-0.8cm} 
\begin{eqnarray}
\nonumber   \text{over} \hspace{0.1cm} \{\rho_{\tau}\} \hspace{5.4cm}
\end{eqnarray}
\vspace{-0.8cm}
\begin{eqnarray}
\nonumber   \hspace{-0.5cm} \text{ and constraints:}\hspace{1cm}  \rho_{\tau} \geq 0, \hspace{0.6cm}\rho_{\tau} ^{T_1}\geq 0.
\end{eqnarray}
A problem with these characteristics is a Semidefinite Programming (SDP) problem which is numerically solvable \cite{activation2012, SDP1996}. We have solved it by using MATLAB \cite{MATLAB} with the YALMIP toolbox \cite{YALMIP} and the solvers SDPT3 \cite{SDPT3} and SeDuMi \cite{sedumi}. We next analyse all these properties on specialised examples.

%%%%%%%%%%%%%%%%%%%%%%%%%
%%%%%%%%%%%%%%%%%%%%%%%%%
\section{Results: Some Quantum Properties of States of Interest}
%%%%%%%%%%%%%%%%%%%%%%%%%
%%%%%%%%%%%%%%%%%%%%%%%%%

In this section, we shall use the formalism and tools already described in the previous sections in order to analyse the following quantum properties of some specific states: entanglement, nonlocality, $k$-copy nonlocality, hidden nonlocality, activation T\&LF, and locality. From now on, we will deal with quantum states in terms of a parameter $p$, i. e., $\rho=\rho(p)$, with the following notation:  if $p>p_{E}$ then the state is entangled, if $p>p_{NL}$ the state is nonlocal, if $p>p_{SA}$ the state is useful for teleportation (an therefore $k$-copy nonlocal), if $p> p_{HN}$ the state contains hidden nonlocality, if $p>p_{T\&LF}$ the program described in the previous section has found an ancillary state that helps to the activation T\&LF, and  if $p\leq p_{L}$ the state is local. We will analyse these properties onto the Werner states reproducing the values reported in \cite{activation2012}. Additionally, we report new activation regions for Isotropic and Hirsch states. 

%%%%%%%%%%%%%%%%%%%
\subsection{Werner-Isotropic (WI) States:}

\begin{figure}[h!]
 \centerline{\includegraphics[height=2.4in,width=3.5in]{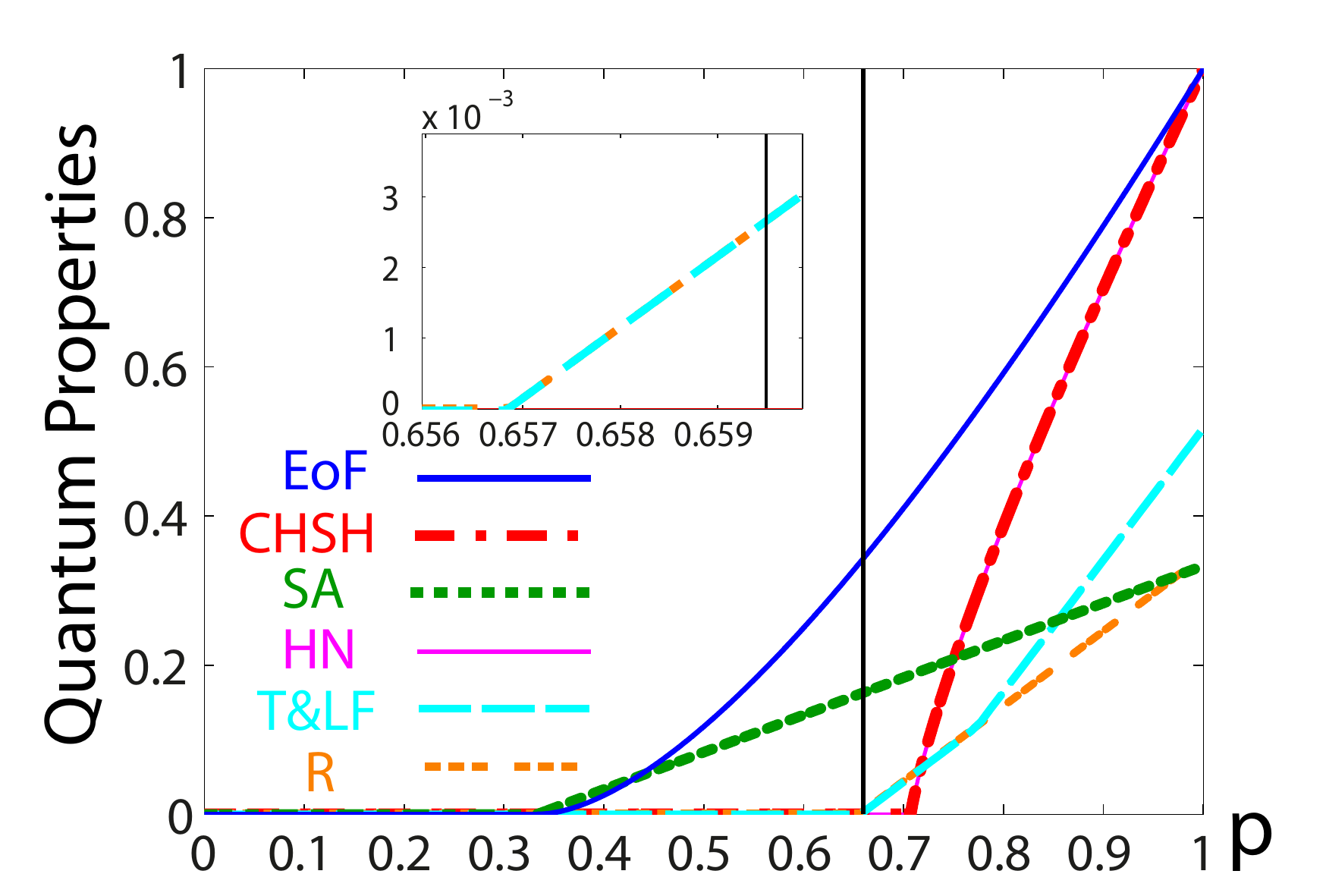}}
\label{fig:fig4}
\caption{Quantum properties for the WI states (\autoref{wistates}). Entanglement-EoF (blue solid thick curve) \autoref{EoF22}. Superactivation-SA (green dotted curve) \autoref{SA22}. Nonlocality-CHSH (red dashed-dotted curve) \autoref{CHSH22}. Hidden Nonlocality - HN (magenta solid thin curve) \autoref{HN22}. Activation T\&LF through the minimisation procedure (\autoref{minimisation}) (cyan dashed curve) for which we have plotted $-5\sigma\left[\tau_{WI}(p)\right]$ in order to make it visible amongst the other properties. R (orange triply dashed curve), activation T\&LF with the ancillary matrix (\autoref{ancillary}), which was obtained by the minimisation procedure (\autoref{minimisation}) at  $p_A=$ 0.6569, and Projective-Locality (black vertical solid line) at $p_{L}=$ 0.6595 according to the best known bound derived in Ref. \protect\cite{Acin2006}.}
\end{figure}

We first consider the so-called Werner states \cite{Werner1989}. Particularly, the two-qubit version of them, in which we refer to them as the Werner-Isotropic (WI) states, and read
\begin{eqnarray}
 \tau_{WI}(p) = p\left| \psi_- \right> \left< \psi_- \right | + \frac{(1-p)}{4}\mathds 1, \label{wistates} 
\end{eqnarray}
\begin{eqnarray}
\nonumber 0\leq p \leq 1, \hspace{0.7cm} \left| \psi_- \right>:=\frac{1}{\sqrt 2} \left (\left| 01 \right>-\left| 10 \right>\right ).
\end{eqnarray}
In \autoref{fig:fig4}, we have plotted the nonlocality-related properties discussed throughout the paper for these WI states, namely,  EoF, CHSH, SA, HN and activation T\&LF. Regarding activation T\&LF, we have plotted $-5\sigma[\tau_{WI}(p)]$ in order to make it visible amongst the other measures. We were also able to extract the ancillary matrix for $p=$ 0.6569 \autoref{ancillary}, which turns out to be useful for all the 0.6569 $<p<1$ region, as we show in the $R(p)$ function plotted in \autoref{fig:fig4}. Therefore, $\tau_{WI}(p) \otimes \rho_{\tau}$ is CHSH-nonlocal after a LF with, 
\begin{eqnarray}
 \rho_{\tau}= \frac{1}{16} \sum_{i,j=0}^3 R_{ij} \sigma_i \otimes \sigma_i \otimes \sigma_j \otimes \sigma_j,
\label{ancillary}
\end{eqnarray}
\begin{eqnarray}
\nonumber \text{where, }\hspace{0.2cm} R:= \frac{1}{9} \begin{pmatrix}9& 3 & 3 & 3\\ 1& -1& 3& -1 \\ 1& -1& 3& -1 \\ 1& -1&3& -1 \end{pmatrix}.
\end{eqnarray}
To complete the results reported in \autoref{fig:fig4}, the following \autoref{tab:tab1} gives the limit bounds of the nonlocality-related properties beyond entanglement for the WI states. A few things should be pointed out: First, even though the nonlocality limit point obtained from the CHSH inequality is $p_{NL}=\frac{1}{\sqrt{2}} \approx$ 0.7071, there exists a slightly better bound which reads $p_{NL}=$ 0.7054 \cite{Hua2015}. Second, CHSH-nonlocality coincides with hidden nonlocality because the WI states are already in a Bell-diagonal form (see \cite{ref30} for a detailed explanation). We see that for the states in the entangled local region $0.3333<p<0.6596$, even though they do not present hidden nonlocality, they present superactivation of nonlocality and activation T\&LF.

\begin{table}[h!]
\center
%\hspace{-0.5cm}\resizebox{1\columnwidth}{!}{%
\begin{tabular}[t]{|c|c|c|c|c|c|}
\hline
 $ p_{E}$& $p_{SA} $ & $p_{T\&LF}$ & $ p_{L}$& $p_{HN}$ & $p_{NL}$ \\ 
\hline
  0.3333 & 0.3333 & 0.6569 & 0.6595& 0.7054 & 0.7054 \\
\hline
\end{tabular}
\caption{Limit values for the nonlocality-related properties beyond entanglement for WI States (\autoref{wistates}).}
\label{tab:tab1}
\end{table}

%%%%%%%%%%%%%
\subsection{Werner States:}

\begin{figure}[h!]
 \centerline{\includegraphics[height=2.5in,width=3.5in]{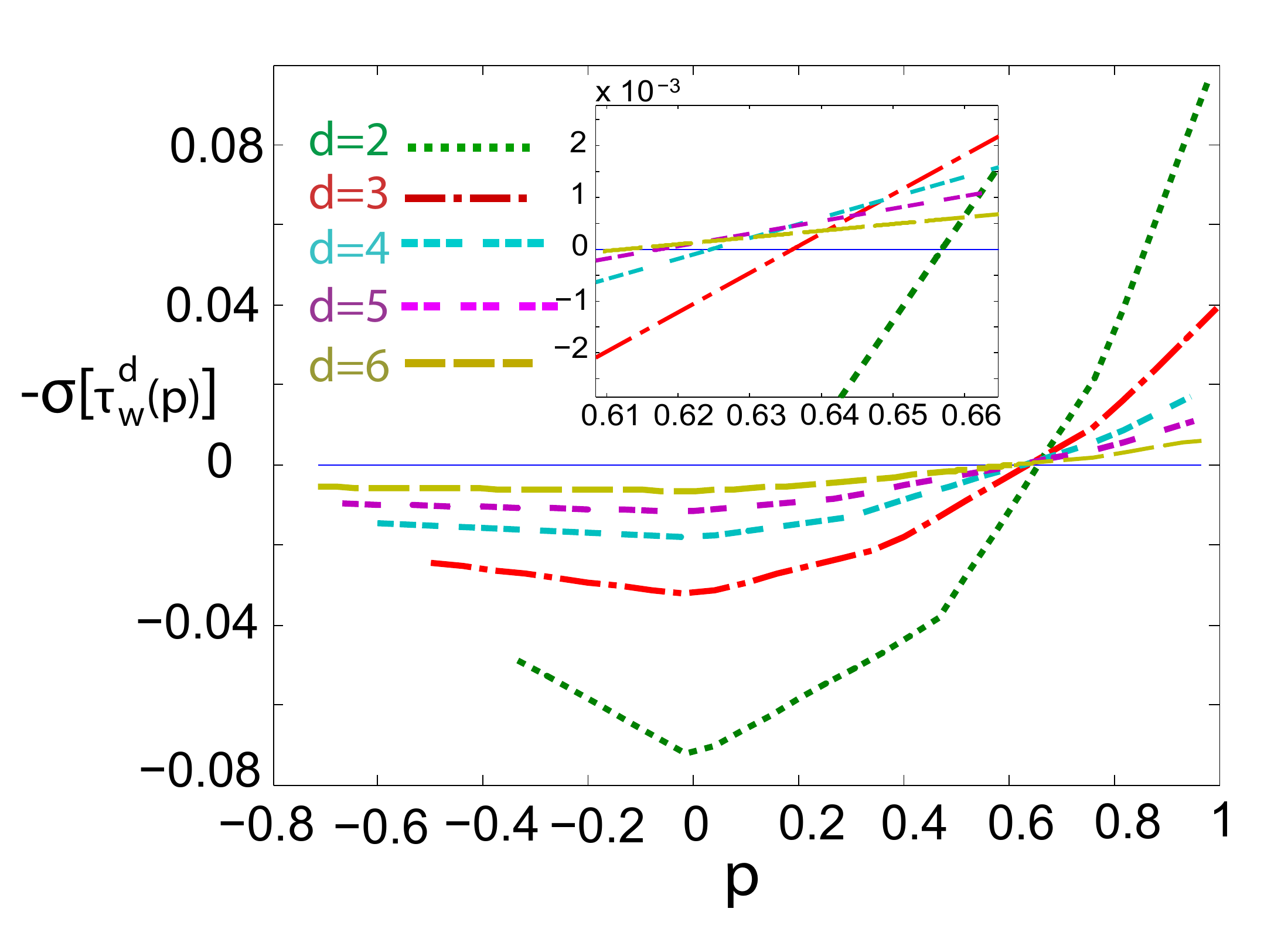}}
\label{fig:fig5}
\caption{Activation T\&NL for Werner states. Minimisation procedure (\autoref{minimisation}) of the function $\sigma[\tau_{W}^d(p)]$ vs parameter $p$ for $\tau_{W}^d(p)$,  high dimensional Werner states (\autoref{wernerstates}). $d=2$ (green dotted curve), $d=3$ (red dahsed-dotted curve), $d=4$ (cyan triply dashed curve), $d=5$ (magenta doubly dashed curve), $d=6$ (yellow dashed curve). The inset shows a zoom of the functions close to the zero value. We reproduce the values reported in \protect\cite{activation2012}, which we report in \autoref{tab:tab2}.}
\end{figure}

We next address the so-called two-qudit Werner states \cite{Werner1989}:
\begin{eqnarray}
 \tau_{W}^d(p) = \frac{ p}{d(d-1)} 2P_{{\rm{anti}}}+ \frac{(1-p)}{d^2}\mathds 1, 
 \label{wernerstates}
\end{eqnarray}
\begin{eqnarray}
\nonumber   \hspace{1.8cm} 1 - \frac{2d}{d+1}\leq p\leq 1, 
\end{eqnarray}
\begin{eqnarray}
\nonumber  P_{{\rm{anti}}}:=\frac{1}{2}\left( \mathds{1}-\sum_{ij}^d  \left| i \right>\left< j \right| \otimes  \left| j \right>\left< i \right|\right).
\end{eqnarray}
In \autoref{fig:fig5}, we have plotted the activation T\&LF by means of the minimisation of the $\sigma( \tau_{W}^d )$ function, according to \autoref{minimisation}) and up to qudits of dimension $d=6$. In \autoref{tab:tab2}, we report the limit values for the nonlocality-related properties discussed throughout the paper, also up to $d=6$. The entanglement and locality limits come from \cite{Werner1989}. The superactivation limit  by means of the FoT comes from \cite{FEFwi2010}, from which it turns out they (as soon as $d>2$) are not useful for teleportation. Therefore, it is unknown if they are superactivable or not, and so, we have marked them in the table with an X. The activation T\&LF column follows the plots shown in the \autoref{fig:fig5} and also reported in \cite{activation2012}. Finally, in the last column, hidden nonlocality from \cite{Popescu1995}. In Appendix A, we give a more detailed description of these bounds.
\begin{table}[h!]
\center
%\hspace{-0.4cm}\resizebox{1\columnwidth}{!}{%
\begin{tabular}[t]{|c |c |c|c|c|c|}
\hline
\textbf {$d$} & $ p_{E}$& $p_{SA} $ & $p_{T\&LF}$ & $ p_{L}$ & $p_{HN}$ \\
\hline
 $2$ &  0.3333 & 0.3333 & 0.6569 & 0.6595 & 0.7054 \\
\hline
 $3$ &  0.2500&X&  0.6360 &0.6667&0.7630 \\
\hline
$4$ & 0.2000 &X&  0.6247 & 0.7500 & 0.7837  \\
\hline
$5$ &  0.1429  &X&  0.6174 & 0.8000 & 0.7944\\
\hline
$6$ &  0.1667 &X&  0.6127 & 0.8333 & 0.8009  \\
\hline %$\sim$
\end{tabular}
\caption{Limit values for nonlocality-related properties beyond entanglement for Werner states (\autoref{wernerstates}) up to $d=6$.}
\label{tab:tab2}
\end{table}

%%%%%%%%%%%%%
\subsection{Isotropic States:}

\begin{figure}[h!]
 \centerline{\includegraphics[height=2.7in,width=3.7in]{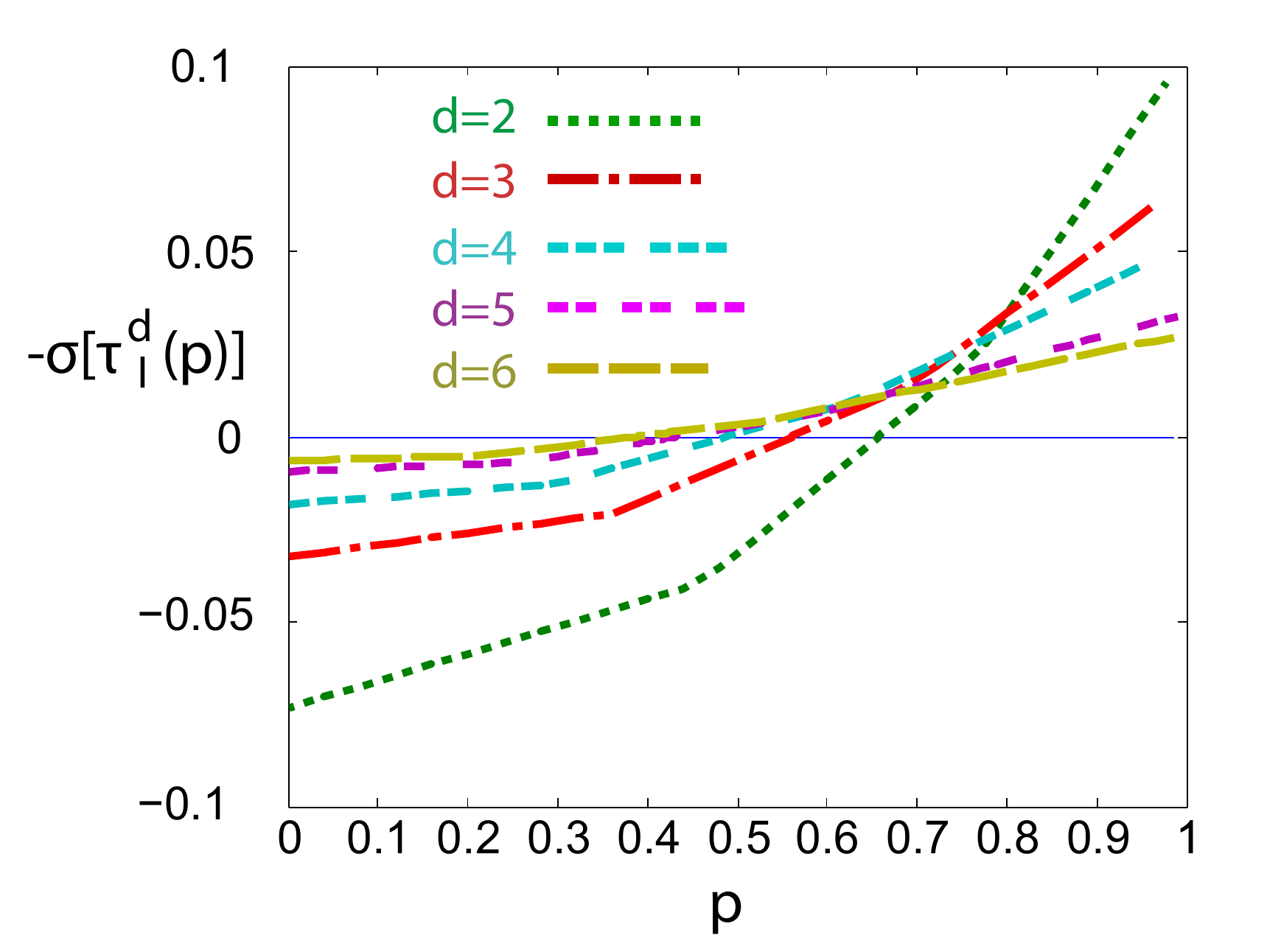}}
\caption{Activation T\&NL for Isotropic states. Minimisation procedure (\autoref{minimisation}) of the function $\sigma[\tau_{I}^d(p)]$ vs parameter $p$ for $\tau_{I}^d(p)$  high dimensional Isotropic states (\autoref{isotropicstates}). $d=2$ (green dotted curve), $d=3$ (red dashed-dotted curve), $d=4$ (cyan triply dashed curve), $d=5$ (magenta doubly dashed curve), $d=6$ (yellow dashed curve). We report in \autoref{tab:tab3} the values we are interested in.}
\label{fig:fig6}
\end{figure}

We now consider the Isotropic states \cite{isotropic1999}:
\begin{eqnarray}
 \tau_{I}^d(p) = p\left| \psi_d \right> \left< \psi_d \right | + \frac{(1-p)}{d^2}\mathds 1, 
  \label{isotropicstates}
\end{eqnarray}
\begin{eqnarray}
\nonumber  0\leq p \leq1, \hspace{0.7cm}  \left| \psi_d \right>:=\frac{1}{\sqrt{d}}\sum_{i=1}^d \left| i i \right>.
\end{eqnarray}
In \autoref{fig:fig6}, we have plotted the activation T\&LF by means of the minimisation of the $\sigma(\tau_{I}^d)$ function, according to \autoref{minimisation} and up to qudits of dimension $d=6$. In \autoref{tab:tab3}, we report the limit values for the nonlocality-related properties discussed throughout the paper up to $d=6$. The entanglement and locality limits are taken from \cite{isotropic1999}. The superactivation limit is obtained from \cite{FEFwi2010} (all of them are useful). Activation T\&LF comes from plots shown in \autoref{fig:fig6}. Finally, nonlocality limit values are obtained from the Collins-Gisin--Linden-Massar-Popescu (CGLMP) inequalities \cite{CGLMP2002}. In Appendix B, we give a more detailed description of these bounds.
\begin{table}[h!]
\center
%\hspace{-0.4cm} \resizebox{1\columnwidth}{!}{%
\begin{tabular}[t]{|c|c|c|c|c|c|}
\hline
\textbf {$d$} & $ p_{E}$& $p_{SA} $ & $p_{T\&LF}$ & $ p_{L}$ & $p_{NL}$ \\
\hline
$2$&  0.3333 & 0.3333 & 0.6569 & 0.6595 & 0.7054 \\
\hline
 $3$&  0.2500& 0.2500 &  0.5606 & 0.4167 & 0.6961\\
\hline
$4$& 0.2000 & 0.2000 &  0.4890 & 0.3611 & 0.6905  \\
\hline
$5$&  0.1429  & 0.1429 &  0.4337 &0.3208 & 0.6872   \\
\hline
$6$&  0.1667 & 0.1667 &  0.3895 & 0.2900 & 0.6849  \\
\hline %$\sim$
\end{tabular}
\caption{Limit values for nonlocality related properties beyond entanglement for Isotropic states (\autoref{isotropicstates}) up to $d=6$.}
\label{tab:tab3}
\end{table}

%%%%%%%%%%%%
\subsection{Hirsch States:}

\begin{figure}[h!]
 \centerline{\includegraphics[height=4.6in,width=3.3in]{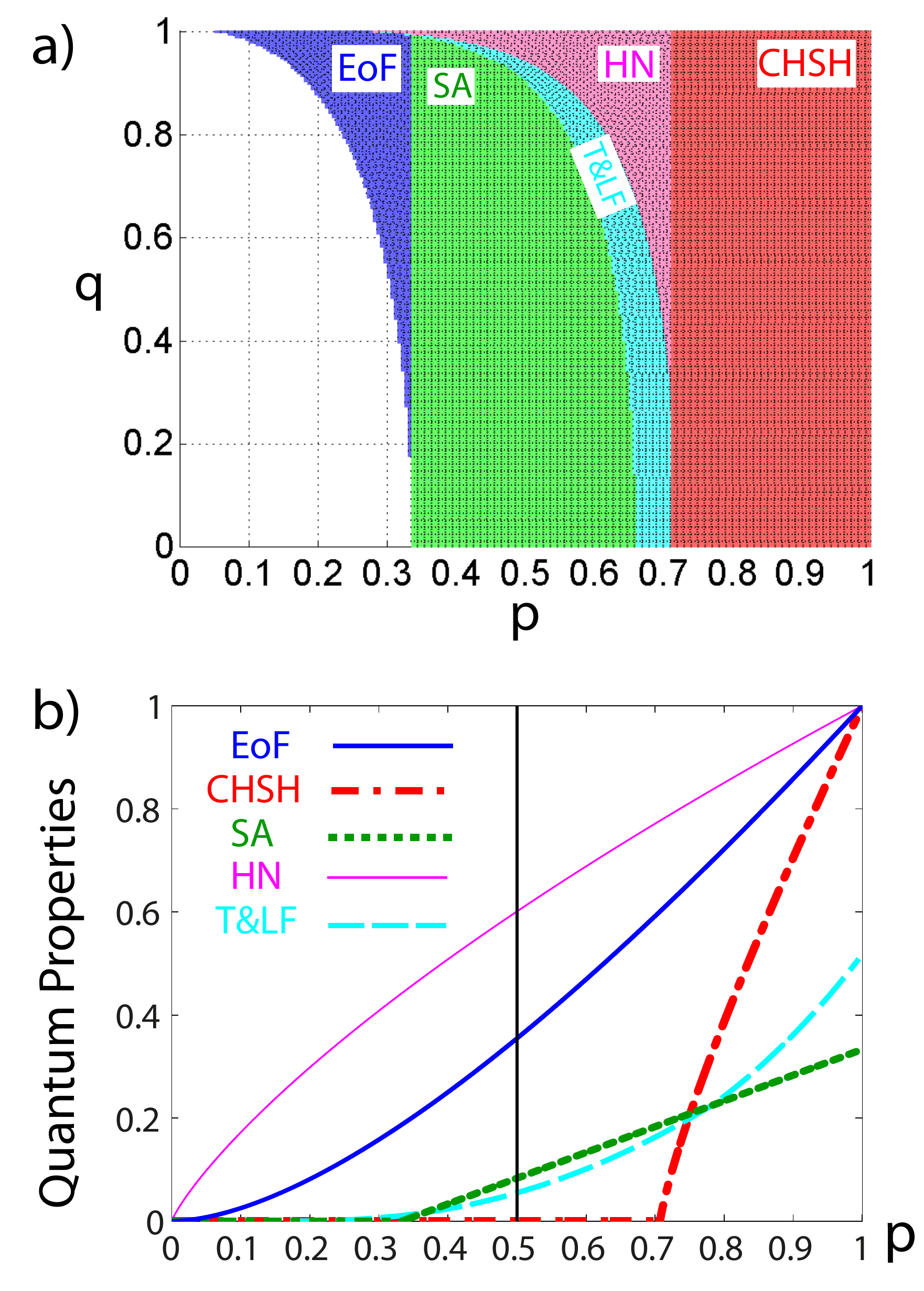}}
\label{fig:fig7}
\caption{{\bf{(a)}} Quantum properties for the two-parameter Hirsch states (\autoref{twoparameter}) with $\sigma=\left| 0 \right>\left< 0 \right |$. Entanglement-EoF (blue) \autoref{EoF22}.  Nonlocality-CHSH (red) \autoref{CHSH22}. Superactivation-SA (green) \autoref{SA22}. Hidden Nonlocality-HN (magenta) \autoref{HN22}. Activation T\&LF (cyan) \autoref{minimisation}. {\bf{(b)}} Quantum properties for one-parameter Hirsch states \autoref{oneparameter} (or two-parameter Hirsch states with $\sigma=\left| 0 \right>\left< 0 \right |$ and $q=1$, previous plot {\bf{(a)}} at $q=1$). Entanglement-EoF (blue solid thick curve) \autoref{EoF22}. Nonlocality-CHSH (red dashed-dotted curve) \autoref{CHSH22}. Superactivation-SA (green dotted curve) \autoref{SA22}. Hidden Nonlocality-HN (magenta solid thin curve) \autoref{HN22}. Activation T\&LF for which we have plotted $-5\sigma\left[\tau_F(p,q)\right]$ in order to make it visible amongst the other properties (cyan dashed curve) \autoref{minimisation}. Projective-Locality (black vertical solid line).}
\end{figure}

Finally, we analyse the  two-qubit states studied by Hirsch, Quintino, Bowles and Brunner \cite{Hirsch2013} which for simplicity we will call Hirsch states:
\begin{eqnarray}
\tau_{F}(p,q,\sigma) = p\left| \psi_- \right> \left< \psi_- \right |  + [1-p]\left[q\sigma+(1-q)\frac{\mathds 1}{2} \right]\otimes \frac{\mathds 1}{2}, 
\label{twoparameter}
\end{eqnarray}
\begin{eqnarray}
\nonumber    0\leq p \leq 1,\hspace{0.5cm} 0\leq q \leq 1, \hspace{0.5cm}\left| \psi \right>:=\frac{1}{\sqrt{2}}(\left|01\right>-\left|10\right>),
\end{eqnarray}
and $\sigma$ and arbitrary one-qubit state. These states can also be thought as a generalisation of the two-qubit WI states (\autoref{wistates}); in fact, we recover them by putting $\sigma=\frac{\mathds 1}{2}$, and $q=1$. In \autoref{fig:fig7} (a), we have plotted the nonlocality-related properties for the two-parameter Hirsch states (\autoref{twoparameter}), using $\sigma=\left| 0 \right>\left< 0 \right |$.  This plot should be understood as the projection of these properties upon the $p-q$ plane. The white region is composed by separable states whilst the rest stands for entangled states and the nonlocality properties that lie within. These properties have been superposed following the hierarchy we expect. Interestingly, these states for $\sigma=\left| 0 \right>\left< 0 \right|$, and $q=1$ become
\begin{eqnarray}
 \tau_{F}(p) = p\left| \psi_- \right> \left< \psi_- \right | + (1-p) \left| 0 \right> \left< 0 \right |\otimes \frac{\mathds 1}{2},
\label{oneparameter}
\end{eqnarray}
for which the authors of Ref. \cite{Hirsch2013}) were able to prove locality, i. e., to build a local model, for all $p>1/2$. In \autoref{fig:fig7} (b), we have plotted the aforementioned properties for the states in \autoref{oneparameter}. In \autoref{tab:tab4}, we have reported the limit values for the aforementioned properties for the states in \autoref{oneparameter}.

%\hspace{-0.3cm}\resizebox{0.96\columnwidth}{!}{%
\begin{table}[h!]
\center
\begin{tabular}[t]{|c|c|c|c|c|c|}
\hline
 $ p_{E}$& $p_{HN} $ & $p_{T\&LF}$  & $p_{SA}$ & $ p_{L}$&$p_{NL}$ \\
\hline
0 & 0 &  0.1716 &0.333& 0.5000 & 0.7071\\
\hline
\end{tabular}
\caption{Limit values for the nonlocality-related properties beyond entanglement for the one-parameter Hirsch states (\autoref{oneparameter}).}
\label{tab:tab4}
\end{table}

%%%%%%%%%
%%%%%%%%%
\section{Discussion}
%%%%%%%%%
%%%%%%%%%

In this work, we have dealt with $6$ properties of quantum states, namely, entanglement, nonlocality, locality, and three generalisations of nonlocality  by means of the activation scenarios: hidden nonlocality HN (or activation through local filtering), $k$-copy nonlocality (or superactivation of nonlocality SA), and  activation through tensoring and local filtering (T\&LF). We stress that there are more general setups, for instance, tensoring between different states or quantum networks. However, we have chosen the already discussed ones, because, at least for two-qubit systems, we can numerically enquire for these properties (with the exception of locality).

The three activation scenarios we have worked with share the mechanisms of whether local filtering or tensoring.  Whilst local filters for two qubits, which could be seen as operations that the experimentalists can locally implement, keep the dimension of the system, tensoring increases it. Additionally, both mechanisms take separable states into separable states (and consequently local states). Therefore, they cannot possibly activate nonlocality on separable states. Thus, they could only be useful for entangled local states.

We then have analysed these properties upon particular states of interest, namely, the so-called Werner, Isotropic, and Hirsch states. These particular choices were made because there exist local bounds for these states which let us consider entangled local states. We remark that unlike the others properties, locality cannot be yet approached numerically. Locality can only be investigated by means of the construction of local models which is not an easy task. 

This paper sheds light into mainly two aspects. First, from a practical point of view, it could be seen as a reference guide to calculate nonlocality-related properties for two-qubit systems. In particular, regarding $k$-copy nonlocality, \autoref{fig:fig3} (a,b) specifies the integer number $k$ necessary for the superactivation. We have checked these properties upon the well known Werner states \autoref{fig:fig4}, \autoref{fig:fig5} and \autoref{tab:tab1}, \autoref{tab:tab2}. Second, we have reported new bounds for other states of interest (namely, Isotropic and Hirsch states). We have chosen these particular states, in the same vein as Werner states, because of their known bounds regarding locality. Even though nonlocality- related properties have already been reported for these states, activation T\&LF has not been calculated yet. We have reported these bounds, filling this gap. Before going into the details of our findings, a note on the activation T\&LF. 

The main theorem regarding activation T\&LF (\autoref{TandLF}) guarantees that all entangled states can be activated in this way. Therefore, from a purely theoretical point of view, it could be seen as pointless to further study this scenario in relation with entanglement. However, the theorem in question does not provide us the matrix necessary for the activation,  which is an important issue from a practical point of view. The numerical approach (\autoref{numerical}) could give us the ancillary matrix in question, and it is open how to find the local filter that maximises the activation. We now proceed to discuss our findings for the Isotropic and Hirsch states.
 
For the so-called two-qudit Isotropic states, even though nonlocality-related properties have been studied, no bounds for activation T\&LF have been reported yet. In \autoref{fig:fig6}, we have calculated these bounds which we report in \autoref{tab:tab3} column 4 ($p_{T\&LF}$) in terms of the dimension $d$ of the qudits. From these results, one can observe the following. Unlike the Werner states' bounds, these Isotropic states' bounds do not cover the local states (given $d>2$, values in \autoref{tab:tab3} column 4 ($p_{T\&LF}$) are still greater than values in \autoref{tab:tab3} column 5 ($p_L$)). However, these new bounds now extend the known nonlocality region (given $d>2$, values in \autoref{tab:tab3} column 6 ($p_{NL}$) are greater than values in \autoref{tab:tab3} column 4 ($p_{T\&LF}$)). Unfortunately, there is no two-qudit characterisation of the set $P_{CHSH}$, unlike two-qubit systems \cite{PG2015}, so we cannot say anything in this regard.

For the so-called two-qubit Hirsch states, even though nonlocality-related properties have been reported, neither bounds for hidden nonlocality nor activation T\&LF have been reported yet. In \autoref{fig:fig7} (a), we have calculated these bounds. First, we remark the hierarchy amongst these properties. All of these properties are inside entanglement and all of them cover the standard definition of nonlocality. Second, we are able to report a $P_{CHSH}$ activation  region for these qubits (which could also be called tensorial activation of hidden nonlocality), here depicted by the cyan region (activation T\&LF) that is not covered by the magenta region (HN) in \autoref{fig:fig7} (a). 

We have depicted the particular case of \autoref{fig:fig7} (a) for $q=1$ in \autoref{fig:fig7} (b) and \autoref{tab:tab4}, because there is a locality bound for these states. First, the states within the locality region ($p<p_L=0.5$) are usually considered useless for quantum protocols based on nonlocality. However, they are now displaying a nice variety of generalised nonlocality-related properties. Actually, all of the three generalisations we are considering here, unlike Werner states in \autoref{fig:fig4} which present SA (all of the local region) and Act T\&LF (a small region) but no HN. Second, comparing this again  with the Werner states in \autoref{fig:fig4}, we are numerically showing that there is not a trivial relation between these generalisations of nonlocality. In particular, there are Werner states with SA but no HN whilst  there are Hirsch states with HN but no SA.

%%%%%%%%%%%%%%
%%%%%%%%%%%%%%
\section{Concluding Remarks}
%%%%%%%%%%%%%%
%%%%%%%%%%%%%%

First, we have reviewed the so far known activation of nonlocality scenarios. We payed particular attention to the hidden nonlocality, $k$-copy nonlocality, and the activation through tensoring and local filtering, in particular, upon two-qubit systems. We have reviewed the numerical approaches required in order to establish a quantification of such quantum properties. For the particular case of two-qudit Werner states, we have reproduced the limit points of the above-mentioned properties.

Second, using the above tools we analysed the activation of nonlocality related properties now for Isotropic and Hirsch states.  In particular, we reported limit points on the activation of nonlocality through tensoring and local filtering that, to the best of our knowledge, have not been reported so far. Additionally, due to the recent result in \cite{PG2015}, we  have calculated hidden nonlocality for two-qubit Hirsch states which has led us to report tensorial activation of hidden nonlocality. 

%%%%%%%%%%%%%
%%%%%%%%%%%%%
\section*{Acknowledgements} 
%%%%%%%%%%%%%
%%%%%%%%%%%%%

We gratefully acknowledge support from COLCIENCIAS Program ``Jovenes Investigadores e Innovadores Virginia Gutierrez de Pineda''  under contract No. RC-0618-2013, from COLCIENCIAS contract No. 71003, Universidad del Valle-internal project No. 7930, the Colombian Science, Technology and Innovation Fund-General Royalties System (Fondo CTeI-SGR) contract No.~BPIN 2013000100007, and from CIBioFi.

%%%%%%%%%%%%%
%%%%%%%%%%%%%
\section*{Conflict of interest} 
%%%%%%%%%%%%%
%%%%%%%%%%%%%

There are no conflicts of interest with funding sources or institutions.

%%%%%
\appendix

%%%%%%%%%%%%%%%%%
%%%%%%%%%%%%%%%%%
\section{Appendix A: Werner States}
%%%%%%%%%%%%%%%%%
%%%%%%%%%%%%%%%%%

Here we show the nonlocality-related properties limit points (except activation T\&LF) for the two-qudit Werner states (\autoref{wernerstates}). They are entangled for $p>p_E$ and local for $p\leq p_L$ with \cite{Werner1989}:
\begin{eqnarray}
\nonumber p_{E} =\frac{1}{d+1},  \hspace{0.5cm}   p_L =\frac{d-1}{d}.
\end{eqnarray}
The explicit calculation of the fidelity of teleportation reported in \cite{FEFwi2010}, shows that all entangled Werner states are not useful for teleportation, i. e., $f_d(\rho_{\rm{ent}})<\frac{1}{d}$. Their hidden nonlocality or nonlocality through LF (only) is checked as in \cite{Popescu1995}, which we detail in what follows. Applying upon the two-qudit Werner states (\autoref{wernerstates}) the local filtering operation given by $P\otimes Q$, with:
\begin{eqnarray}
   P =|0\left >\right<0|_A+|1\left >\right<1|_A,\hspace{0.5cm}
  Q =|0\left >\right<0|_B+|1\left >\right<1|_B,
  \label{filters}
\end{eqnarray}
we obtain the two-qubit state $\rho(p):= [P\otimes Q]\rho_{W}^d(p)$,
\begin{eqnarray}
\nonumber   \rho(p)=\frac{ p}{d(d-1)} 2|\psi\left >\right<\psi|+ \frac{(1-p)}{d^2}\mathds 1_2 \otimes \mathds 1_2,
\end{eqnarray}
with $\left | \psi \right>:=\frac{1}{\sqrt{2}}\left(\left | 01\right>-\left | 10\right> \right )$. Normalizing we obtain:
\begin{eqnarray}
\nonumber   \rho(p)&=&\left[\frac{d(d-1)d^2}{2pd^2+(1-p)4d(d-1)}\right] \\
\nonumber  &&\times\left[ \frac{ p}{d(d-1)} 2|\psi \left >\right<\psi|+ \frac{(1-p)}{d^2}\mathds 1_2 \otimes \mathds 1_2 \right].
\end{eqnarray}
Now, checking its CHSH maximal violation by means of the criterion in \autoref{criterion}, we have that the second part vanishes, whilst the first part achieves maximum violation, i. e., $2\sqrt{2}$, then:
\begin{eqnarray}
\nonumber   \rm{max} B_{\rho (p)}=\left[\frac{d(d-1)d^2}{2pd^2+(1-p)4d(d-1)}\right] \\ \nonumber  \times \left[\frac{p}{d(d-1)} 2(2\sqrt 2) \right].
\end{eqnarray}
We have CHSH violation when $\rm{max} B_{\rho_{W}^d}(p)>2$ (\autoref{CHSH}). After some algebra we obtain that this holds for $p\geq p_{NL}$, with:
\begin{eqnarray}
\nonumber   p_{NL} = \frac{4(d-1)}{2d(\sqrt 2-1)+4(d-1)}.
\end{eqnarray}
For instance, the values we are interested in:
\begin{itemize}
\item $d=3$, $\hspace{0.5cm} p_{NL} = \frac{4}{17}(3\sqrt 2-1) \approx$ 0.7630.
\item $d=4$, $\hspace{0.5cm} p_{NL} = \frac{3}{7}(2\sqrt 2-1) \approx$ 0.7837.
\item $d=5$, $\hspace{0.5cm} p_{NL} = \frac{8}{41}(5\sqrt 2-3) \approx$ 0.7944.
\item $d=6$, $\hspace{0.5cm} p_{NL} = \frac{5}{14}(3\sqrt 2-2) \approx$ 0.8009.
\end{itemize}
Which are the values reported in \autoref{tab:tab2} and in \cite{activation2012}. In \cite{activation2012} they also reported that, after numerical optimisations over possible filters, filters given by \autoref{filters} are optimal, in the sense that, the obtained two-qubit system violates CHSH at its best.

%%%%%%%%%%%%%%%%%
%%%%%%%%%%%%%%%%%
\section{Appendix B: Isotropic States}
%%%%%%%%%%%%%%%%%
%%%%%%%%%%%%%%%%%

Here, we show the nonlocality-related properties limit points (except activation T\&LF) for the two-qudit Isotropic states (\autoref{isotropicstates}). They are entangled for $p>p_E$ and local for $p\leq p_L$, with \cite{isotropic1999}:
\begin{eqnarray}
\nonumber p_E =\frac{1}{d+1}, \hspace{0.5cm} p_L =  \frac{-1+\sum_{k=1}^d\frac{1}{k}  }{d-1}.
\end{eqnarray}
The explicit calculation of the fidelity of teleportation reported in \cite{FEFwi2010}, shows that all entangled Isotropic states are useful for teleportation, i. e., $f_d(\rho_{\rm{ent}})>\frac{1}{d}$. Finally, nonlocality is checked through the violation of the CGLMP inequalities \cite{CGLMP2002}. 

%%%%%%%%%%
%%%%%%%%%%
\section*{References}
%%%%%%%%%%
%%%%%%%%%%

\bibliographystyle{unsrt}
\bibliography{bibliography.bib}

\end{document}